\documentclass[12pt,preprint]{emulateapj}

\usepackage{amssymb}
\usepackage{amsmath}

\def\aj{\rm{AJ}}                    
\def\apj{\rm{ApJ}}                 
\def\apjl{\rm{ApJ}}                
\def\apjs{\rm{ApJS}}                        
\def\mnras{\rm{MNRAS}}   

\usepackage{epsfig}
\shorttitle{SDSS 0956+5128: Extreme Velocity Offsets}
\shortauthors{Steinhardt, C. L. et al.}

\begin{document}


\title{SDSS 0956+5128: A Broad-line Quasar with Extreme Velocity Offsets}


\author{Charles L. Steinhardt}
\affil{Kavli IPMU, University of Tokyo, Kashiwanoha 5-1-5, Kashiwa-shi, Chiba, Japan}

\author{Malte Schramm}
\affil{Kavli IPMU, University of Tokyo, Kashiwanoha 5-1-5, Kashiwa-shi, Chiba, Japan}

\author{John D. Silverman}
\affil{Kavli IPMU, University of Tokyo, Kashiwanoha 5-1-5, Kashiwa-shi, Chiba, Japan}

\author{Rachael Alexandroff}
\affil{Department of Astrophysical Sciences, Princeton University, Ivy Lane, Princeton, NJ 08544, USA}
\author{Peter Capak}
\affil{California Institute of Technology, MC 249-17, 1200 East
California Boulevard, Pasadena, CA 91125, USA}
\author{Francesca Civano}
\affil{Harvard-Smithsonian Center for Astrophysics, 60 Garden St., Cambridge, MA 02138, USA}
\author{Martin Elvis}
\affil{Harvard-Smithsonian Center for Astrophysics, 60 Garden St., Cambridge, MA 02138, USA}
\author{Dan Masters}
\affil{University of California, Department of Physics and Astronomy, Riverside, CA 92508, USA, Observatories of the Carnegie Institution of Washington, Pasadena, CA, 91101}
\author{Bahram Mobasher}
\affil{University of California, Department of Physics and Astronomy, Riverside, CA 92508, USA}
\author{Petchara Pattarakijwanich}
\affil{Department of Astrophysical Sciences, Princeton University, Ivy Lane, Princeton, NJ 08544, USA}
\author{Michael A. Strauss}
\affil{Department of Astrophysical Sciences, Princeton University, Ivy Lane, Princeton, NJ 08544, USA}


\begin{abstract}
We report on the discovery of a Type 1 quasar, SDSS 0956+5128, with a surprising combination of extreme velocity offsets.  SDSS 0956+5128 is a broad-lined quasar exhibiting emission lines at three substantially different redshifts: a systemic redshift of $z \sim 0.714$ based on narrow emission lines, a broad Mg{\small II} emission line centered 1200 km/s bluer than the systemic velocity, at $z \sim 0.707$, and broad H$\alpha$ and H$\beta$ emission lines centered at $z \sim 0.690$.  The Balmer line peaks are 4100 km/s bluer than the systemic redshift.  There are no previously known objects with such an extreme difference between broad Mg{\small II} and broad Balmer emission.  The two most promising explanations are either an extreme disk emitter or a high-velocity black hole recoil.  However, neither explanation appears able to explain all of the observed features of SDSS 0956+5128, so the object may provide a challenge to our general understanding of quasar physics.
\end{abstract}

\keywords{black hole physics --- galaxies: evolution --- galaxies: nuclei --- quasars: general --- accretion, accretion disks}


\section{Introduction}

The broad emission lines distinctive of quasar spectra are believed to come from high-velocity gas in the ``broad-line region" (BLR), $\sim 0.1-1$ pc from the central engine, with typical full width at half maximum (FWHM) in the range of 2000-20000 km/s (cf. \citet{Petersonbook}).  Prominent broad emission lines visible in optical spectra at $z \sim 0.7$ include H$\beta$ and Mg{\small II}, with H$\alpha$ visible in the near infrared and C{\small IV} in the UV.  Quasar broad emission lines are usually assumed to be predominantly virial, and are typically symmetric \citep{Murray1995}.  Indeed, black hole mass estimators have been developed predicated upon this virial assumption and Gaussian line profiles \citep{McLure2002,McLure2004,Vestergaard2002,Vestergaard2006,Wang2009,Onken2008,Shen2008,Risaliti2009,Rafiee2011}.  

However, there are many known sources of non-virial motion, often leading to asymmetric broad line profiles that may have peaks offset from the systemic redshift of the host.  Strong inflows or outflows would shift the peak of BLR lines.  Similarly, radiation pressure and/or quasar winds can yield an asymmetric C{\small IV} line \citep{Marconi2009, Marziani2010}, as can broad absorption lines.  Typical asymmetries are several hundreds to 2000 km/s.

A much larger offset in the broad emission line peaks may be caused by eccentric disk emission.  Emission lines originating in a relativistic, eccentric disk may have very asymmetric profiles with double peaks \citep{Chen1989,Eracleous1995}.  Numerous examples of these disk or double-peaked emitters have been found in the Sloan Digital Sky Survey (SDSS) quasar catalog \citep{Strateva2003}, typically based upon H$\beta$ and H$\alpha$ profiles and Mg{\small II} for objects at higher redshift \citep{Luo2009}.  The blue peak of these Balmer lines is often offset by several thousand km/s from the wavelength corresponding to the systemic narrow-line redshift \citep{Stratevathesis}.  If the asymmetry from Doppler boosting is strong enough, there may appear to be only one, blueshifted peak in a skewed line profile (e.g., \citet{Eracleous1995} Fig. 4c).

The post-merger recoil of central supermassive black holes can also give rise to offset broad emission-line peaks.  If each galaxy involved in a merger has already formed a supermassive black hole, dynamical friction brings the two black holes into a black hole binary \citep{Yu2002,DiMatteo2005,Escala2004,Merritt2005,Mayer2006,Dotti2006,Colpi2009}, from which the binary pair can eventually coalesce into one supermassive black hole (cf. \citet{Pretorius2007}).  Numerical simulations reveal a final phase in which the emission of gravitational radiation produces a spin-dependent recoil of the resulting black hole, ranging from $\lesssim 200$ km/s for non-spinning black holes to perhaps $\gtrsim 1000$ km/s for spinning black holes in special orientations \citep{Campanelli2007,Herrmann2007,Schnittman2007,Loeb2007}.  Observational signatures of such a recoil might include a quasar broad-line region moving at $\gtrsim 1000$ km/s relative to its host galaxy and the quasar narrow lines \citep{Loeb2007}.

There are no known, confirmed examples of recoiling supermassive black holes.  Several examples of binary black holes have been found at low redshift ($z < 0.05$) \citep{Komossa2003,Ballo2004,Hudson2006,Rodriguez2006,Bianchi2008,Comerford2009,Liu2010,McGurk2011,Max2007,Barrows2012}, with separations in the kiloparsec range.  Attempts to systematically search for recoiling black hole candidates in which optical spectral show multiple line systems at slightly different redshifts \citep{Bonning2007,Boroson2009,Jahnke2009,Liu2010} have yielded several candidates \citep{Komossa2008}, but there is typically a more mundane, alternative explanation as well \citep{Shields2009}.  In particular, HE 0450-2958 \citep{Magain2005,Hoffman2006} was first claimed to have no host galaxy and thus be a good candidate for an ejected black hole, but it may have a faint host galaxy after all, and even shows signs of ongoing star formation, while it does not show the necessary offset between narrow and broad emission lines that must be characteristic of a recoiling black hole \citep{Merritt2006,Kim2007,Jahnke2009}.  Most recently, CID-42 \citep{Comerford2009,Civano2010}, at $z = 0.359$, includes two optical sources offset by $1200$ km/s lying 2.4 kpc apart in the same galaxy, and appears to be a good candidate for either a recoiling black hole or a system containing two AGN, one Type 1 and the other Type 2.  New observations using Chandra appear to rule out the latter scenario \citep{Civano2012}.

In this work, we describe SDSS J095632.49+512823.92 (hereafter SDSS 0956+5128), a broad-lined Type 1 quasar exhibiting lines with three substantially different redshifts: a systemic redshift for quasar narrow emission lines, a second, 1200 km/s bluer, associated with broad Mg{\small II} emission, and a third, 4100 km/s bluer than the narrow lines, for broad H$\alpha$ and H$\beta$ (Table \ref{table:linesummary}).  This combination of spectral lines is very difficult to understand.  Explanations for a blueshifted broad-line region relative to the host include (1) an undetected host galaxy lying at a lower redshift along the same line of sight, (2) a disk or double-peaked emitter \citep{Strateva2003}, (3) a multiple-black-hole system in the host galaxy, or (4) a recoiling black hole.  However, we know of no explanation for a Type I quasar in which one broad line is blueshifted by 2900 km/s relative to the other.  No other systems with three such velocities were found in a search of the SDSS DR8 \citep{Aihara2011} quasar catalog.  Therefore, SDSS 0956+5128 appears to be a rare object, and thus might be difficult to fit into existing models because it may be a relatively short-lived state.  If so, SDSS 0956+5128 may reveal a previously unobserved phase in the SMBH life cycle.

In \S~\ref{sec:object}, we describe SDSS observations of SDSS 0956+5128.  \S~\ref{sec:confirming} describes followup spectroscopic observations of H$\alpha$, H$\beta$, and Mg{\small II} 4-5 years (in the rest frame) after the initial SDSS spectrum, confirming that there are indeed three different velocities in this system.  Subaru IRCS J, H, and K band images of the host galaxy, in addition to GALEX and 2MASS observations, enable us to address whether a recent merger has occurrred. (\S~\ref{sec:host}).  Finally, in \S~\ref{sec:discussion} we consider the two models that might be most capable of producing such a system.  We discuss whether SDSS 0956+5128 might be an extreme disk emitter with unique features.  Alternatively, we consider the possibility that SDSS 0956+5128 might be a recoiling post-merger supermassive black hole.  Neither appears to be a good explanation for the three observed velocities in SDSS 0956+5128.

\section{SDSS 0956+5128 in SDSS}
\label{sec:object}

In a 2002 DR3 spectrum \citep{SDSSDR3}, SDSS 0956+5128 (Fig. \ref{fig:segue}) clearly shows a broad-lined Type I quasar, selected initially based upon its colors in five filters \citep{Richards2002} for spectroscopic followup. 
\begin{figure}[!ht]
\epsfxsize=3.25in\epsfbox{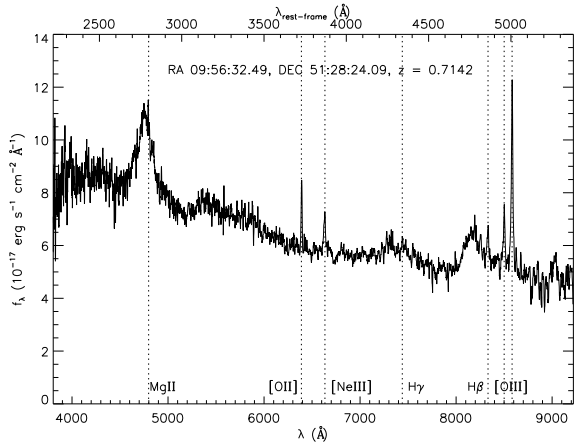}
\caption{Spectrum of SDSS 0956+5128 from SEGUE-2.  The identified redshift of $z = 0.714$ is a good fit to narrow emission lines such as [O{\small II}], [O{\small III}], and the narrow component of H$\beta$.  However, the broad emission lines H$\beta$ and Mg{\small II} are offset by 4100 km/s and 1200 km/s, respectively.}
\label{fig:casspec}
\end{figure}
 The SDSS redshift of $z = 0.714$ is a good fit for the narrow emission lines, including [O{\small II}]$\lambda 3727$, [O{\small III}]$\lambda 4959,5007$, and the narrow component of H$\beta$ (Fig. \ref{fig:casspec}).  The broad emission lines, however, lie at lower velocities.  We fit the continuum, Fe{\small II} emission, and the broad Mg{\small II} line following \citet{McLure2004} and \citet{Shen2008}.  The best-fit Mg{\small II} line has a redshift $0.7071 \pm 0.0006$  and FWHM $12800 \pm 490$ km/s, with the $11 \sigma$ difference providing compelling evidence for a BLR offset velocity of 1200 km/s from the narrow lines.  The $z = 0.7071$ fit has a $\chi^2/\textrm{DOF}$ of 0.97 over wavelengths within 10900 km/s (2$\sigma$) of the best-fit line centroid, compared to 1.36 for the best $z = 0.714$ fit.

The broad component of H$\beta$ is also blueshifted relative to $z = 0.714$.  The centroid of the broad component is at $z = 0.690$, apparently blueshifted by approximately 4100 km/s relative to the narrow lines, compared to 1200 km/s for Mg{\small II}.  We also note that Mg{\small II} is symmetric, while H$\beta$ is not, even though they have similar ionization potentials and should be emitted from gas clouds at a similar radius from the central black hole (cf., \citet{McLure2002}). 


\begin{table*}[!ht]
\begin{center}
\caption{Emission Line properties of SDSS 0956+5128.}
\begin{tabular}{|c|c|c|c|}
\hline 
Line & z & Observations & Clear asymmetry \\
\hline 
Mg{\small II} $\lambda$ 2798 & 0.707 & DR3, SEGUE-2 & No \\
$[\textrm{O {\small II}}] \lambda$ 3727 & 0.714 & DR3, SEGUE-2 & No \\
H$\beta$ $\lambda$ 4861 & 0.690 & DR3, SEGUE-2, DEIMOS & Yes \\ 
$[\textrm{O {\small III}}] \lambda$ 4959 & 0.714 & DR3, SEGUE-2, DEIMOS & No \\
$[\textrm{O{\small III}}] \lambda$ 5007 & 0.714 & DR3, SEGUE-2, DEIMOS & No \\ 
H$\alpha$ $\lambda$ 6563 & 0.690 & TripleSpec & Yes \\
\hline  
\end{tabular}
\label{table:linesummary}
\end{center}
\end{table*}

\subsection{Confirming Observations of Broad Emission Line Velocity Offsets With Optical Spectra}
\label{sec:confirming}

\begin{figure}[!ht]
 \epsfxsize=3.25in\epsfbox{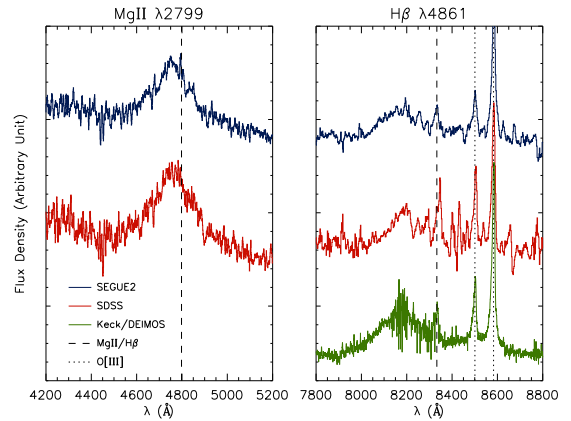}
\caption{Comparison of Mg{\small II} and H$\beta$ as observed by SEGUE-2 (dark blue), Keck/DEIMOS (green), and in the original SDSS spectrum (red).  THe Keck/DEIMOS spectrum has the highest signal-to-noise, followed by SEGUE-2 and then SDSS DR3.  The expected wavelength given the redshift of the systemic narrow lines is indicated (dashed).  The broad emission line centroids lie at similar velocities in all spectra, and the spectra are consistent with being identical.}
\label{fig:segue}
\end{figure}

 The offset between the broad H$\beta$ and Mg{\small II} emission lines is sufficiently surprising and difficult to explain that we must consider whether measurement error, fitting error, or absorption from the host galaxy might be responsible for mis-locating one of the line centroids.  SDSS 0956+5128 was observed a second time using the SDSS spectrographs as part of SEGUE-2 \citep{Yanny2009} (Fig. \ref{fig:casspec},\ref{fig:segue}), seven years after the initial spectrum (it was erroneously targeted as a candidate BHB star).  The SEGUE-2 spectrum has higher signal-to-noise and is consistent with the previous SDSS measurement of Mg{\small II} at $z = 0.707$, H$\beta$ at $z = 0.690$, and the narrow lines at $z = 0.714$.

Because the H$\beta$ profile is asymmetric and has such a large offset, we performed two additional followup observations (Table \ref{table:observations}).  Using Keck/DEIMOS \citep{Faber2003}, a higher-resolution and higher signal-to-noise spectrum using the 830G grating ($R \sim 3200$, compared to $\sim 2000$ for SDSS) was taken on November 23-24, 2011 under photometric conditions and 0.6-0.8$^{\prime\prime}$ seeing.  On November 23rd two 600s exposures were taken at a slit PA of 90 and on November 24th 900s and 300s exposures were taken at a slit PA of 0.   The 1200G grating tilted to 7800~\AA\ was used with a OG550 blocking filter and 0.5" slit, resulting in a spectral resolution of 0.43~\AA\ FWHM measured from sky lines.  The data were reduced with a modified version of the DEIMOS reduction package which accounts for guider drift, variable seeing, wavelength calibration errors, and non-photometric conditions. 

The spectro-photometric standard star HZ-44 was used for relative flux calibration.  Regions of telluric absorption and strong spectral lines in HZ-44 were masked, and then a 10th order polynomial was fit to the ratio of measured to expected flux.  
After scaling out the baseline throughput, gaussian lines were fit to the residual flux ratio at the location of known telluric absorption lines to account for atmospheric absorption.  Both spectra show H$\beta$ in the same location as the SDSS spectra (Fig. \ref{fig:segue}).  The H$\beta$ profile shows similar asymmetry in both SDSS and Keck spectra. 

\begin{table*}[!ht]
\begin{center}
\caption{Summary of observations of SDSS 0956+5128.}
\begin{tabular}{|c|c|c|c|}
\hline 
Instrument/Survey & MJD & Wavelengths & Exposure Time (s) \\
\hline 
SDSS plate 902 & 52409 & 3900 -- 9200 \AA & 2700 \\
SEGUE-2 plate 3320 & 54912 & 3900 -- 9200 \AA & 900 \\
DEIMOS & 55888-55889 & 6500 -- 9000 \AA & 2400 \\
TripleSpec & 55922 & 9500 -- 24600 \AA & 1920 \\
Subaru IRCS & 55972 & H-band & 2736 \\
Subaru IRCS & 55972 & J-band & 3168 \\
Subaru IRCS & 55972 & K$_p$-band & 3195 \\
Subaru IRCS & 55984 & K-band & 2910 \\
Subaru IRCS & 55985 & K-band & 2500 \\
\hline  
\end{tabular}
\label{table:observations}
\end{center}
\end{table*}

\subsection{Near-infrared Spectroscopy}

\begin{figure}[!ht]
\begin{center}
 \epsfxsize=3.25in\epsfbox{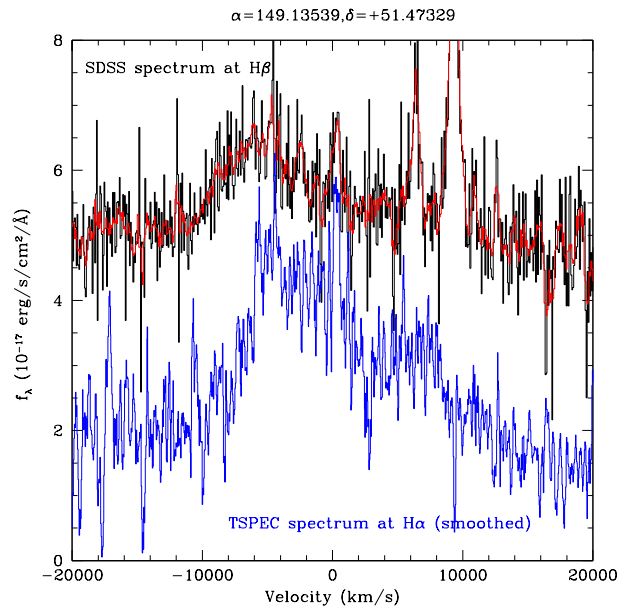}
\end{center}
\caption{Comparison of H$\alpha$ as observed by TripleSpec on the ARC 3.5m and H$\beta$, as observed by SEGUE-2.  The broad and narrow components are similarly offset and the lines have similar overall profiles. Velocities are indicated relative to the best-fit $z \sim 0.714$ for the narrow emission lines.  The apparent absorption feature at $\sim 2500$ km/s in H$\alpha$ coincides with atmospheric absorption lines.  While the Triplespec spectrum has been corrected for atmospheric absorption from observations of a nearby A star, this feature may be a residual rather than absorption associated with the quasar or its host.}
\label{fig:halpha}
\end{figure}

In addition, an infrared spectrum was taken using TripleSpec \citep{Wilson2004,Rayner2003,Cushing2004,Vacca2003} (R $\sim 2800$) for eight 240s exposures in December 2011 on the Astrophysical Research Consortium 3.5m at Apache Point Observatory.  This spectrum reveals a broad H$\alpha$ emission line at $z = 0.690$ with a profile comparable to H$\beta$ (Fig. \ref{fig:halpha}), with no evidence of strong absorption lines in either line.  We note that although the strong asymmetry in H$\alpha$ and H$\beta$ is one characteristic of disk or double-peaked emitters \citep{Eracleous1995,Strateva2003}, as above, the large offset between H$\beta$ and Mg{\small II} is not characteristic of the few examples with well-measured H$\beta$ and Mg{\small II}.  We conclude that the measured centroids for the Balmer lines indeed lie at a different redshift than the narrow lines.  In Table \ref{table:observations}, we summarize the observations, both spectroscopy and imaging, of SDSS 0956+5128 that we use in this study.

\section{The Host Galaxy}
\label{sec:host}

A key question in understanding SDSS 0956+5128 lies in determining whether there are multiple galaxies along the same line of sight.  If there are multiple systems, at least the offset between broad emission lines and narrower galactic lines would be easily explained.  However, if SDSS 0956+5128 is indeed one system with three distinct velocities, it would be more puzzling.  In particular, if the peculiar features of SDSS 0956+5128 are due to a black hole recoil, we can make several predictions about its host.  Most importantly, the black hole should indeed be associated with the $z \sim 0.714$ galactic counterpart.  Attenuation of the quasar SED due to a dusty host that recently underwent a merger could be evidence that we are investigating one system, not two.

\subsection{Broad-band SED}

\begin{figure}[!ht]
\begin{center}
 \epsfxsize=3.25in\epsfbox{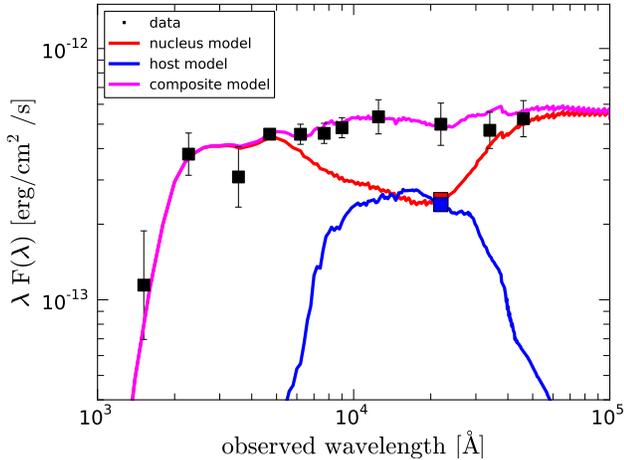}
\end{center}
\caption{Best-fit decomposition of the observed SDSS 0956+5128 SED into an active nucleus (red) and host galaxy (blue).  The sum of the two components (magenta) is matched to the observed SED (black)}
\label{fig:decomposition}
\end{figure}

We consider the broad-band SED of the source combining SDSS data with GALEX \citep{GALEX}, WISE \citep{WISE}, and 2MASS \citep{2MASS}.  We find that the optical data require A$_{\rm V}$=0.3 reddening of the SDSS composite quasar spectrum \citep{VandenBerk2001,Richards2006}.  Even this is not enough to reproduce the observed level of IR emission.  Therefore, we have added a galaxy spectrum with primarily an old stellar population from \citet{Maraston2005}.   We make use of the K-band (\S~\ref{subsec:kband}) image decomposition to constrain the AGN template and the galaxy template in our SED fit. This is important to reduce the degeneracies between the different models.  This decomposition is shown in Fig. \ref{fig:decomposition}.  We note that some photometric data points are likely to be affected by strong emission lines (i.e., Ly$\alpha$, H$\alpha$).  Both the quasar and inferred host galaxy appear to be dusty.

From our decomposition, we roughly estimate the stellar mass of the galaxy to be $\sim 8 \times10^{10} M_{\odot}$.  We also use our Mg{\small II} and the strength of the nearby continuum fit to infer a central black hole mass of $\log (M/M_{\odot}) = 8.65$ using the Mg{\small II}-based virial mass estimator calibrated by \citet{McLure2004}.  This estimate may be incorrect if there is substantial nonvirial motion, although the Mg{\small II} line appears symmetric despite its offset.  The estimated black hole mass and galaxy luminosity are consistent with a central black hole-galaxy pair lying near the $z=0$ $M_{BH} - L_{\textrm{gal}}$ relation \citep{Bentz2009}.  However, our inferred galaxy is likely a factor of $\sim 4$ lower mass than necessary to lie on the $M_{BH} - M_{bulge}$ relation \citep{Haering2004}.  Because the inferred host lies either on or below these relations, any different counterpart corresponding to the quasar should be nearly as bright as our inferred host.

\subsection{High-resolution IR Imaging with Subaru}
\label{subsec:kband}

\begin{figure}[!ht]
\begin{center}
 \epsfxsize=3.25in\epsfbox{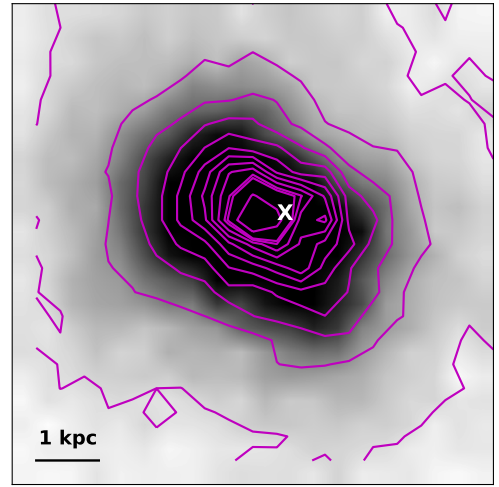}
\end{center}
\caption{Infrared emission of the host galaxy of SDSS 0956+5128 as observed by Subaru IRCS (K-band), with a resolution of 0.15 arcsec.  The quasar emission, with its centroid marked in white, has been subtracted.  The quasar point source appears to lie away from the center of an asymmetric galaxy, which may indicate recent merger activity.  Contours are marked on a linear scale.}
\label{fig:subaru}
\end{figure}

Stronger evidence that the quasar lies in the inferred host comes from followup J, H and K band images using the Subaru Infrared Camera and Spectrograph (IRCS) with adaptive optics (AO) \citep{Hayano2008,Hayano2010}.  We performed this imaging using the AO188 Adaptive Optics system in Laser Guide Star (LGS) mode, with the LGS equidistant from the quasar and PSF star in order to improve PSF stability.  The K-band image with 52mas pixels shown in Fig. \ref{fig:subaru} was taken with a net exposure time of 66 min but varying conditions.  The PSF was then modeled using a star of brightness similar to the main quasar point source, as a combination of a Gaussian core and Moffat wing, and rotated to correct for mismatched alignment between the source, LGS, and PSF star.  

Only the K-band image shown in Fig. \ref{fig:subaru}, with a PSF of 0.15 arcsec, was suitable for our purposes.  The second K-band image was taken with a 20mas pixel scale.  Therefore, the flux per pixel is much lower and our integration time was too short.  We did not reach the neccessary depth to determine properties of the host, although extended host emission can be detected. The J,H,Kp images were taken in service mode with a PA=90°. This leads to different distances from the guide star to the PSF and the QSO. Anisoplanatism degrades the seeing and cannot be controlled with effectively one PSF star.  So, we cannot predict the shape of the PSF at the quasar position.  Therefore, we have decided to reject those images in the analysis.

We then decompose the K-band image into a point source and host using GALFIT v. 3.0 \citep{Peng2010}.  First a PSF-only model is fit to the data, then a Sersic profile is fit for the host galaxy contribution.  We find that the AGN-host flux ratio is $1.0 \pm 0.4$, and the best-fit parameters for the host are those of a compact, early-type galaxy ($n_{\textrm{Sersic}} \sim 3.7-4.5, a/b=0.50 \pm 0.15, R_{1/2}\sim 1.3\textrm{kpc}$).  The resulting picture is thus consistent with a point source lying in a single host galaxy (Fig. \ref{fig:subaru}).

There appear to be asymmetric structures in the central pixels, but due to the difficulties in controlling the AO PSF, we cannot be certain that they are real.  Surprisingly, the quasar appears not to be located at the highest concentration of starlight, likely the galactic bulge.  If SDSS 0956+5128 does not include a recoiling black hole, this is difficult to understand even in the case of a merger, particularly coupled with the high velocity.  However, the apparent offset may be an artifact caused by imperfections in the point-source subtraction.  

\section{Discussion}
\label{sec:discussion}

SDSS 0956+5128 appears to be unique in the SDSS DR8 catalog in that there are emission lines at three velocities in the system, corresponding to $z \sim 0.714$ (narrow lines), $z \sim 0.707$ (Mg{\small II}), and $z \sim 0.690$ (H$\alpha$ and H$\beta$).  The combination of these lines is startling.  We considered four possible explanations: (1) an undetected host galaxy lying at a lower redshift along the same line of sight, (2) a disk or double-peaked emitter \citep{Strateva2003}, (3) a multiple-black-hole system in the host galaxy, or (4) a recoiling black hole.  

If there were two Type 1 quasars involved, there should be two Mg{\small II} and two H$\beta$ broad lines, lying 2900 km/s apart.  SDSS 0956+5128 exhibits just one broad emission line of each species, which would require two unusual quasars, one with very weak Mg{\small II} and another with very weak H$\beta$, both lying at a different velocity from that of the narrow lines.  Even if a third supermassive black hole were responsible for the narrow lines, a triple system model has the same flaw, that each black hole should have a discernible set of lines, and SDSS 0956+5128 only has one set of narrow lines, one Mg{\small II} line, and one set of Balmer lines.

Disk emitters have been observed with offsets of 4100 km/s or more between broad and narrow lines \citep{Strateva2003}, but none of the handful of disk emitters with spectra containing both H$\beta$ and Mg{\small II} has a clear offset between H$\beta$ and Mg{\small II}.  Given their large velocity offsets, disk emitters are likely examples of extreme accretion physics, so perhaps SDSS 0956+5128 is an even more extreme system than previous examples, producing sharply different line profiles for lines even of similar ionization potential.  Similarly, a recoiling black hole could potentially have a large offset between various broad and narrow lines, but none has ever been observed.  We describe the case for each interpretation below.

\subsection{Is SDSS 0956+5128 an Extreme Disk Emitter?}
\label{subsec:diskemitter}

The Balmer lines in this object are strongly skewed with a strong peak blueward of the systemic velocity.  Such objects cannot be fit with a typical, circular disk, even with Doppler boosting.  However, they can be fit by models such as an eccentric disk \citep{Eracleous1995} in which an additional parameter is added.


Sequences of various line profiles with different parameters can be found in \citet{Eracleous1995} and \citet{Stratevathesis}.  A line profile with a strong blue peak 4000 km/s bluer than the central value and a low-amplitude red peak is predicted for models with high values of the disk orientation $\varphi_0$, high eccentricity, and and a small inner pericenter distance $\xi_1$.  We note that our ability to fit such a lineshape using an eccentric disk may not be all that surprising due to the large number of parameters.  The seven parameters (in addition to the central black hole mass) allowed by an eccentric disk, combined with noise and the difficulties of decomposing the spectrum into H$\beta$ broad and narrow components and [O{\small III}]$\lambda 4959,5007$, might mean such a model will fit almost any skewed lineshape.  Further, other sorts of non-circular disk may also be consistent with this line profile, i.e., such a fit is not unique.

Regardless, an eccentric disk model does not produce different line shapes for Mg{\small II} and H$\beta$, since the ionization potentials of the two species are similar.  The parameters producing a strongly skewed H$\alpha$ and H$\beta$ will produce a similarly-shaped Mg{\small II} line, also with a peak at $z \sim 0.690$ instead of $z \sim 0.707$.  So, although the Balmer lines alone could be consistent with SDSS 0956+5128 as a disk emitter, Mg{\small II} is inconsistent with this interpretation.  

\subsection{Is SDSS 0956+5128 a Recoiling Black Hole?}
\label{subsec:recoil}

\citet{Merritt2006} proposed that a post-merger, recoiling black hole could produce a symmetric BLR, offset by $~\gtrsim 1000$ km/s from the NLR velocity.  Could SDSS 0956+5128 be an example of such a recoil?  We must show that the supermassive black hole is associated with the galaxy emitting the narrow lines at $z \sim 0.714$.  Because of the high redshift, we cannot rule out that the black hole lies in another galaxy along the line of sight.  Rather, we must rely on indirect evidence, considering whether the observed galaxy is consistent with being the host of the SMBH responsible for Balmer and Mg{\small II} emission.

The Subaru IRCS imaging reveals only one host, indicating that if there are multiple galaxies along the same line of sight, most of the galactic flux is coming from one, largest galaxy.  The observed galaxy is a promising recoil candidate in that a decomposition into a galactic and quasar SED shows extinction that could be characteristic of a dusty galaxy.  Mergers would likely be associated with strong extinction, although galaxies not undergoing mergers can also be dusty at many stages of their evolution.  

Moreover, the inferred host galaxy mass is $8 \times 10^{10} M_{\odot}$, with a luminosity close to that expected for a $10^{8.65} M_{\odot}$ black hole lying on the $M - L_{\textrm{gal}}$ relation.  A $10^{8.65} M_{\odot}$ central black hole is unlikely to lie in a substantially less luminous galaxy than the one detected by SDSS, and therefore the broad-line region is unlikely to belong to a quasar with a different galactic counterpart.  

The presence of broad emission lines is evidence that if this is a recoiling black hole, the recoil took place in the recent past.  Using an $\alpha$-disk \citep{Shakura1973} model, the accretion disk can temporarily survive the recoil and the broad line region will continue to emit over a lifetime of \citep{Loeb2007}
\begin{equation}
t_{\textrm{disk}} = 8.4 \times 10^6 \alpha^{-0.8}_{-1}\eta^{0.4}M_7^{1.2}{v_8}^{-2.8} \textrm{ yr}.
\end{equation}
For a coalesced black hole mass of $10^{8.65} M_{\odot}$ ($M_7 = 45$) and speed of $v_8 = 1.21$, the dimensionless combination of radiative efficiency and Eddington ratio $\eta = 0.20$ according to the bolometric luminosity given in \citet{Shen2008}.  Assuming a typical viscosity parameter of $\alpha = 0.1$, this yields a lifetime of $t_{\textrm{disk}} \sim 1.4 \times 10^8$ yr for the broad-line region.  The BLR lifetime would be shorter if the Balmer line value of $v_8 = 4.1$ were used.  Thus, we should expect coalescence to have taken place within the past 140 million years.  This is likely comparable to the dynamical timescale for the host galaxy, so if this is indeed a recoiling black hole, followup observations might find evidence of a recent merger.  The quasar-subtracted host galaxy fit would be consistent with a recent merger, but is insufficient to prove one has occurred.  Alternatively, there might have been a long time delay between the merger and coalescence.

During these 140 million years, ignoring gravitational effects from the host galaxy, the ejected black hole could travel a distance of $d \sim v_{\textrm{ej}}t_{\textrm{disk}} \sim 175 kpc$ for $v_{\textrm{ej}} \sim 1200$ km/s, or well outside of the host galaxy.  Depending upon the mass of the host and the current separation, it is possible that SDSS 0956+5128 contains a recoiling black hole that will escape its host.  Followup observations capable of resolving multiple point sources and determining both the distance to the center of the galaxy and mass of the host would be needed in order to determine its eventual fate.

Thus, the Mg{\small II} ($z \sim 0.707$) line by itself, combined with the quasar narrow lines at $z \sim 0.714$ and AO observations of the host, presents a picture consistent with a post-recoil supermassive black hole.  However, the Balmer lines are entirely inconsistent with this explanation; they are better explained as a disk emitter.  Perhaps the explanation is that SDSS 0956+5128 is in some intermediate state in which a recoil has led to an eccentric disk, yet these two scenarios seem incompatible given the similar ionization potentials of H$\beta$ and Mg{\small II}. 

Might these scenarios be compatible?  Because of the number of degrees of freedom involved in disk emitter models, it is possible to fit the Balmer lines of SDSS 0956+5128 as centered at the $z \sim 0.707$ redshift of Mg{\small II} rather than of the narrow lines, again with large $\varphi_0$, high eccentricity, and and a large inner pericenter distance $\xi_1$.  If the impulse delivered to the central black hole at coalescence is non-adiabatic, it may drive a circular accretion disk to become highly eccentric.  There might then be some brief window in which Balmer emission is coming from an eccentric disk but Mg{\small II} is not yet.  One test of this picture would be the C{\small IV} line profile, which cannot be investigated from the ground because it requires UV spectroscopy.  Because C{\small IV} has an ionization potential placing it closer to the central black hole than H$\beta$, it must also be double-peaked.  If such a model is plausible, though, it requires a combination of rare events: (1) a merger; (2) the correct spin and geometry for a large recoil; and (3) timing such that the quasar is observed in the brief time when the Balmer lines have become strongly asymmetric but the slightly higher-radius Mg{\small II} emission is not yet asymmetric.  Since there are no known examples merely where (1) and (2) both occur, this scenario seems unlikely.  However, this object appears to be unique among a catalog of $> 10^4$ quasars containing both H$\beta$ and Mg{\small II}, so it could be a rare glimpse of an object in transition between a circular and eccentric disk.

What is clear, however, is that SDSS 0956+5128 is a challenge for current models of quasar accretion.  If related to a post-coalescence black hole, it would be particularly compelling because it was identified from the spectrum alone, without the ability to resolve the host galaxy, since most mergers occur at too high redshift for us to resolve the host.  An ability to detect such sources from a large survey such as SDSS gives us the opportunity to better understand merger rates and galaxy evolution and to better calibrate our expectations for planned gravitational wave detectors.  If SDSS 0956+5128 is an exotic disk emitter, it will teach us something about the nature of quasar accretion, and in particular that H$\beta$ and Mg{\small II}, despite similar ionization potentials, can arise from quite distinct regions.  And, if SDSS 0956+5128 is in a rare, post-merger transition state, a proper model may allow a better understanding of the quasar life cycle or duty cycle.

The authors would like to thank Steve Bickerton, Kevin Bundy, Masataka Fukugita, Pat Hall, Daniel Proga, Robert Quimby, Allan Sadun, Masayuki Tanaka, and Alexander Tchekhovskoy for valuable comments.  This work was supported by the World Premier International Research Center Initiative (WPI Initiative), MEXT, Japan.  MAS, PP, and RA acknowledge the support of NSF grant AST-0707266.  CLS and ME thank the Aspen Center for Physics and the NSF Grant \#1066293 for hospitality during the editing of this paper.


\bibliographystyle{apj}

\end{document}